\documentclass[conference,twocolumn,11pt]{IEEEtran}

\def\BibTeX{{\rmfamily B\kern-.05em{\scshape i\kern-.025em b}\kern-.08em \TeX}}

\newif\ifpdf
    \ifx\pdfoutput\undefined
    \pdffalse 
\else
    \pdfoutput=1 
    \pdftrue
\fi

\usepackage[latin1]{inputenc}
\usepackage{algorithm}
\usepackage{enumerate}

\ifpdf
    \usepackage[pdftex]{graphicx}
\else
    \usepackage{graphicx}
\fi

\centerfigcaptionstrue

\graphicspath{{.}{fig/}}

\usepackage{amssymb,amsmath}

\begin{document}



\title{Architectural Considerations for a Self-Configuring Routing
Scheme for Spontaneous Networks}

\author{
\authorblockN{Ignacio Alvarez-Hamelin}
\authorblockA{
    LPT~-- CNRS \\
    University of Paris XI \\
    B{\^{a}}timent 210 \\
    91405 Orsay Cedex - France\\
    {\tt Ignacio.Alvarez-Hamelin@lri.fr} 
} \and
\authorblockN{Aline Carneiro Viana}
\authorblockA{
    IRISA~--~INRIA \\
    Campus de Beaulieu \\
    35042 Rennes cedex - France \\
    {\tt aline.viana@irisa.fr}
}
\and
\authorblockN{Marcelo Dias de Amorim}
\authorblockA{
    LIP6~--~CNRS \\
    University of Paris VI \\
    8, rue du Capitaine Scott \\
    75015 Paris - France \\
    {\tt marcelo.amorim@lip6.fr}
}
}

\maketitle
\begin{abstract}

Decoupling the permanent identifier of a node from the node's
topology-dependent address is a promising approach toward completely
scalable self-organizing networks. A group of proposals that have
adopted such an approach use the same structure to: address nodes,
perform routing, and implement location service. In this
way, the consistency of the routing protocol relies on the
coherent sharing of the addressing space among all nodes in the
network.
Such proposals use a logical tree-like structure where routes in this
space correspond to routes in the physical level. The advantage of
tree-like spaces is that it allows for simple address
assignment and management. Nevertheless, it has
low route selection flexibility, which results in low routing
performance and poor resilience to failures. 
In this paper, we propose to
increase the number of paths using incomplete hypercubes.
The design of more complex structures, like
multi-dimensional Cartesian spaces, improves the resilience and
routing performance due to the flexibility in route selection.
We present a framework for using hypercubes to implement indirect 
routing. This framework allows to give a solution adapted to the 
dynamics of the network, providing a proactive and reactive routing
protocols, our major contributions. We show that, contrary to traditional
approaches, our proposal supports more dynamic networks and is more
robust to node failures. 

\end{abstract}


\section{Introduction}
\label{intro}

A scalable location (lookup) service is one of the main design
blocks of a completely self-organizing architecture for
spontaneous networks. In traditional wired networks, location
information can be easily embedded into the topological-dependent
node address, which also uniquely identifies the node in the
network. In self-organizing networks, however, a source only knows
the destination's {\em identifier}, and this identifier does not
give any clue of the destination's {\em address}. There is no
static relation between the node's location and the node's
identifier as a consequence of the spontaneity and adaptability of
the network.

In response to these requirements, distributed hash tables (DHT)
can be adopted as a scalable substrate to provide
location-independent node
identification~\cite{ratnasamy01,stoica.tnet03,rowstron01,blazevic01,jinyangli00,xue01}.
The functionalities of decoupling identification from location,
and of providing a general mapping between them, have made the DHT
abstraction an interesting principle to be integrated at network
layer. Thus, {\em indirect routing} systems that offers a powerful
and flexible rendezvous-based communication
abstraction~\cite{blazevic01,jinyangli00,xue01,benjie02,eriksson04,viana.winet04}
can be implemented.

A number of works have already proposed to use DHTs in routing
protocols. These works can be classified in two main groups, which
differ in the way the DHT structure is
deployed~\cite{viana.adhoc05}. In the first group, the addressing
and the lookup models are completely independent and routing is performed
at the designed addressing structure. A DHT structure is
defined to distribute and locate information among the nodes in the
addressing structure. Examples of proposals in the literature that
implement this approach are:
Terminodes~\cite{blazevic01,hubaux01,terminode},
Grid~\cite{jinyangli00,grid}, and DLM~\cite{xue01}. Most of them
assume, however, that nodes know their geographic coordinates
through some positioning system ({\em e.g.}, GPS). In the second
group classification, the same structure deployed to address nodes 
and consequently to perform routing, is also used by the lookup 
model. This model describes a coherent sharing of the addressing space 
among the nodes, which determines the consistency of the routing protocol.
Tribe~\cite{viana.winet04}, PeerNet~\cite{eriksson04,eriksson03},
Landmark~\cite{tsuchiya88,tsuchiya87}, and L+~\cite{benjie02} are
examples of such protocols.

The proposals that fall in the second group proved that it is
possible to build a logical and mathematical structure from mere
connectivity between nodes. Routing using this mathematical space
gives the exact behavior of the routing mechanism in the physical
layer. Nevertheless, they lack of robustness since their space
sharing mechanism follows a tree structure. Although simple to
implement, a tree offers low flexibility in route selection.
Furthermore, tree structures are not robust to node mobility, since
a node departure causes the breakage of the tree.

Motivated by these observations, in this paper, we propose to use
incomplete hypercubes instead of trees. Contrary to trees,
hypercubes allow the establishment of multiple paths between any
two nodes, which increases the robustness of the topology to
mobility. Indeed, according its literal concept, a tree not allows 
nodes, in its subtree, to be 
connected to nodes in others subtrees. Moreover, a tree is
a $2$-dimensional structure. Otherwise, in a hypercube nodes
can communicate in a $d$-dimensional space, which allows
multiple paths among nodes.
We expect then to represent at least a part of the
broadcast nature of wireless scenarios through the multiple
dimensions of a hypercube. 
In wireless environments, the connectivity is controlled by
the density and communication range of nodes, which can be
relatively large.

Our contributions are twofold. First, we propose a proactive
routing approach, where routes are determined {\em a priori}.
Second, we propose a reactive protocol that establishes routes on
an on demand basis. While the proactive approach is more adapted
to quasi static networks, the reactive protocol is indicated to
mobile networks. We show through a number of examples that our
proposals are promising and are more robust to dynamic networks
than the existent related tree-like approaches.

The remainder of the paper is organized as follows. In
Section~\ref{ind_rout}, we present the indirect routing model
approach with related work and the proposed architecture. We
introduce the hypercube used as addressing space in
Section~\ref{addr_spa}. Section~\ref{pro_reac} presents our
approach and discusses routing-specific issues. Some cases of
study are addressed in Section~\ref{prac_cases_sty} and
Section~\ref{discuss} discusses the applicability and future 
researches of our proposal.

\section{Indirect routing strategy}\label{ind_rout}

The indirect service model is instantiated as a rendezvous-based
communication abstraction. Nodes called {\em rendezvous nodes} are
responsible for storing the location information of others nodes in
the topology. Routing is performed indirectly and the rendezvous
nodes translate a node's identifier into its location-dependent
address in the topology.

\begin{figure}[h]
\begin{center}
\includegraphics[width=3cm]{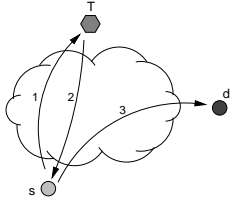}
\end{center}
\caption{Lookup (arrows 1 and 2) and direct communication (arrow 3)
phases in a DHT-based routing procedure} \label{fig:indirect}
\end{figure}

Routing is performed through a topology-dependent technique. Every
node is identified by its position in the topology, which is
translated into a topology-dependent address. It is important to
underline that the only way of routing is by using this address. In
the general case, every node has three identifiers. The first one,
called universal identifier, $U$, is supposed to be known by any
other node that are supposed to communicate with the node. This
identifier is independent of any network-level characteristics. It
can be a word, a numerical value, or even an IP-like address. The
second identifier, the virtual address $V$, is a translation of $U$
into the virtual addressing space, ${\mathcal V}$. This identifier
is used to name the correspondent rendezvous node. The last
identifier, the relative address $E$, is the current
topology-dependent address of the node. Observe that the relative
address changes if the node moves, but both the universal and
virtual identifiers remain unchanged. Fig.~\ref{fig:indirect}
illustrates the steps of the routing procedure and the use of the
described identifiers.

When source $s$ wants to communicate with destination $d$ and has no
idea of $d$'s relative address, it first contacts the node
responsible for storing the relative address of node $d$ (arrow~1).
Call this node $T_d$. Thus, the message sent by $s$ will travel in
the network until it is received by $T_d$. Note that node $s$ does
not know $E_d$, but it knows $V_d$ (obtained from $U_d$). Node $T_d$
knows the relative address $E_d$ because node $d$ has previously
informed $T_d$ about its current address. The rendezvous node $T_d$
plays the role of a ``rendezvous'' point where the location of node
$d$ is stored. The particularity of this approach is that the
rendezvous point is virtually identified and can be any physical
node in the network. Rendezvous nodes are distributed and depend
only on the nodes' identifiers. When contacted by $s$, $T_d$
responds with a message containing the relative address of node $d$,
$E_d$ (arrow~2). Node $s$ can now communicate directly with $d$
(arrow~3).

\subsection{Related work}~\label{ind_rout:rela_wk}

In the traditional Internet model, routing information is embedded into
the topological-dependent node address, {\em i.e.} IP addresses
have been defined for both \emph{identifying} and \emph{locating} a node in the
network. This does not work well in mobile networks (even if they
are not self-organized networks), because permanent node addresses cannot include
dynamic location information, which invalidates topology
information. More recently, a number of flooding-based
protocols have been used to address this problem in the
specific case of ad hoc networks. Nevertheless, it has been
observed that these architectures do not scale well beyond a few
hundred nodes~\cite{broch98,tseng99}. For instance, in sensor or
wireless mesh networks,
where the potential number of addressable nodes may be in the
order of thousands, current solutions cannot be used.

Most proposed routing algorithms for self-organizing networks
distribute the topology information to all nodes in the network.
Thus, following the idea of indirection routing, the {\em
i}3~\cite{stoica02} proposes an overlays-based infrastructure that
offers a rendezvous based communication abstraction. {\em i}3
decouples the act of sending from the act of receiving: sources
send packets to a logical identifier and receives express interest
in packets sent to this identifier. {\em i}3 uses a set of servers
that store identifiers and map packets with these identifiers to
{\em i}3 nodes interested in receiving the packets. This approach
combines the generality of IP-layer solutions with the versatility
of overlay solutions. Our proposition uses a similar concept of
indirect routing, however, it is not based in an overlay
infrastructure and is independent of IP-layer.

L+~\cite{benjie02} proposes an improved version of 
Landmark~\cite{tsuchiya88,tsuchiya87}
routing,  which is better suited to large ad hoc wireless networks.
This protocol describes a more scalable address lookup service and algorithm
improvements that react better to node mobility. An L+ node
updates one location server for each level in the landmark
hierarchy. L+ uses a routing algorithm similar to
DSDV~\cite{perkins94} and keeps more than just the shortest route
to each destination. Nevertheless, L+ and Landmark creates a
tree-based hierarchical topology where nodes are placed, 
offering a low flexibility in route selection.

Tribe~\cite{viana.winet04} is a rendezvous-based routing protocol
for self-organizing networks. By managing regions of a logical
addressing space, Tribe nodes route in a hop-by-hop basis with
small amount of information and communication cost. Nodes that are
physically close in the network also manage close regions in the
Tribe addressing space. Thus, the main component of Tribe is its
proposed simple manageable addressing space used to assign
addresses to nodes. Nevertheless, this space is also a tree-like
structure, which limits paths by the hierarchical structure of a
tree~-- there is only one path between any two nodes.

Similarly to Tribe, PeerNet~\cite{eriksson03} is a peer-to-peer
based network layer for dynamic and large networks. The address
reflects the node's location in the network and is registered with
the respective identifier in the distributed node lookup service.
In PeerNet, the addresses are organized as leaves of a binary
tree~-- the address tree. PeerNet routing is a recursive procedure
descending through the address tree. Thus, in contrast to Tribe,
PeerNet routing disseminates information about the global state of
the network, and nodes maintain a routing table that has $l = \log
N$ entries, {\em i.e.} $O(\log N)$ per-node state (where $N$ is a
number of nodes in the network). Because of the address tree
organization, a node movement may require the assignment of new
addresses to several nodes in PeerNet infrastructure, which
implicitly generates many updates in lookup entries.

\subsection{Increasing the number of paths connections}~\label{ind_rout:dim}

The design of a self-organized network architecture requires an
efficient combination of robustness and complexity. The resilience
of existent proposals and, consequently, the performance of the
routing protocols are strongly related to the complexity of the
deployed addressing structure. On the one hand, tree-like
structures ({\em e.g.}, L+~\cite{benjie02}, Tribe~\cite{viana.winet04},
and PeerNet~\cite{eriksson04}) lead to simple manageable spaces.
Nevertheless, they have low route selection flexibility, which
results in low routing performance and poor resilience to
failures/mobility. Their low complexity is obtained at the cost of
some loss of robustness. On the other hand, more complex
structures, like multidimensional Cartesian spaces, improve the
resilience and routing performance due to the flexibility in route
selection. The associated addressing and location
models, however, become more complex and require a tight association between
the logical and physical planes. In this paper, we propose to 
increase the number of paths connections through hypercubes.


Hypercubes have the inherent property of multiple paths between any
couple of nodes, given a good and interesting logical-topological
mapping. This possibility gives the following improvements.
First, traffic can be well balanced, in contrast to what occurs in
a tree, where the root is heavily charged. This characteristic
allows to use more efficiently the bandwidth. Another important
improvement is that distances in a network are closer to real
distances, which is not necessarily true in a tree. This makes
communications shorter. Finally, a hypercube allows to use
different routing methods thanks to its logical-topological
mapping (proactive and reactive routing), {\em i.e.} the network
could have a routing schema adapted to the dynamics of the
network.

In the following sections, we present our addressing
system and explain how hypercube representation allows the
specification of a logical structure where proactive/reactive routing 
approaches can be exploited while the lookup service is
performed in a simple way.

\section{Address Spaces based on hypercubes}~\label{addr_spa}

In this section, we describe how to implement a virtual addressing
space based on a hypercube structure.

\subsection{A very brief overview of hypercubes}~\label{addr_spa:hyp_ovw}

The hypercube is a generalization of a 3-dimensional cube to an
arbitrary number of dimensions $d$. Each node of the $d$-hypercube
has coordinates {\tt 0} or {\tt 1} for each dimension, covering
all the combinations. This implies that the total number of nodes
is $2^d$. Each node is linked to all nodes whose coordinates
differ only in one dimension. For example, the cube has a node at
coordinates ($0,0,0$), or simply {\tt 000}, which is connected
with nodes at coordinates {\tt 001}, {\tt 010} and {\tt 100},
because all these nodes differ only in one of their dimensions.
Thus, the degree, {\em i.e.} the number of edges, of each
node equals the dimension $d$.

\begin{figure}[h]
\begin{center}
\includegraphics[width=6cm]{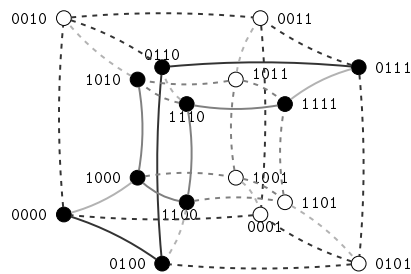}
\end{center}
\caption{Hypercube of dimension $d=4$} \label{routinghyp}
\end{figure}

The most important property of the hypercube is the adjacency of
nodes generated by its construction. Fig.~\ref{routinghyp}
displays a hypercube of dimension 4. We can use the coordinates of
a node as its network address, then the length of the address is $d$. It
easy to see that the distance between two nodes is measured by
XORing the two addresses. For example, the distance between nodes
\texttt{0100} and \texttt{0111} is 2, because there are two
different bits between these nodes, {\em e.g.}, a route could be
\texttt{0100 -> 0110 -> 0111}.

We find interesting examples of hypercube use in: parallel
computing~\cite{Leighton91,Euseuk01}, peer-to-peer
networks~\cite{scholsserAl03}, genetic codes~\cite{jimenez94},
fault-tolerant and redundant systems~\cite{dajin_wang01}, message
stability detection in distributed systems~\cite{friedman99}, parallel 
multiprocessor systems~\cite{slack03}, data communication~\cite{saad85}.

\subsection{The network layer}~\label{addr_spa:net_lay}

\begin{figure}[h]
\hspace{-8mm}\includegraphics[width=99mm]{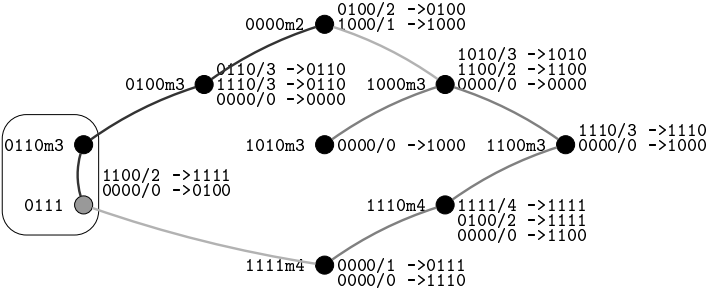}
\caption{Spontaneous network: physical position of nodes}
\label{routingtopo}
\end{figure}

Using node coordinates in the hypercube as its relative address
$E$, it is possible to map a physical network into a logical one.
For an arbitrary physical network, the corresponding mapping
produces an incomplete hypercube, because the number of nodes
present is less than $2^d$, and their physical connection
possibilities do not necessarily correspond to all edges of the
hypercube. We display an arbitrary network in
Fig.~\ref{routingtopo} and its representation on the hypercube in
Fig.~\ref{routinghyp}, where present nodes are filled in black.
Fig.~\ref{routingtopo} also has the routing tables at the right
side of the node, which will be treated latter.

Considering nodes in Fig.~\ref{routingtopo} have a circular
coverage radius, then some nodes use fewer than their possible physical
connections. For example, node {\tt 0100} has a physical
connection with node {\tt 1010}, but their addresses differ in
more than one bit and they are not connected in the hypercube
structure. We say then that the hypercube is incomplete.
Nevertheless, even loosing some connections, the network can take
advantage of the hypercube adjacency for routing.

One way to improve this mapping, {\em i.e.} more physical
connections become edges in the hypercube, is assigning multiple
addresses to some nodes. Since two nodes might not be neighbors in
the hypercube although being physically connected, this allows us
to better represent physical adjacencies.

Summarizing, the information stored in each node is composed of
the main address, the secondary addresses and its addressing
space. The main address corresponds to a network or relative
address $E$, which is given during the connection process. When a
new node joins the network, the main address  is selected by
itself from the addresses proposed by its neighbors (already
connected to the network). Before obtaining the main address, the
new node could chose one or more secondary addresses, if it were
connected to other physical neighbor nodes which are not adjacent
in the hypercube, {\em i.e.} their network addresses $E_i$ are not
adjacent to the new node's main address. For example the node {\tt
0110m3} in Fig.~\ref{routingtopo}, has it main address and the
secondary one: \texttt{0111}. This secondary address is used for
connecting nodes {\tt 0110} and \texttt{1111}, because
\texttt{0111} is adjacent to \texttt{1111}, {\em i.e.} they only
differ in one bit.

Each node manages an addressing space. This addressing space is
used to: $(i)$ store the database for address resolution
queries,\footnote{The rendezvous node stores the $U
\rightarrow E$ entry.} and $(ii)$ give addresses to new nodes. The
later function implies the delegation of a corresponding portion of
addressing space.

The addressing space of a node is determined by its main address
and a mask. This mask is represented by the number of ``ones''
from the left side, {\em e.g.}, \texttt{m3} is the mask
\texttt{1110} because the address length is $d=4$. The address and
its mask (doing bitwise logic AND) gives the addressing space
managed by the node. This method is very similar to IP subnet
masks, because the part with zeros corresponds to the addressing
space managed by the node. For instance, node \texttt{0000m2} in
Fig.~\ref{routingtopo} manages addresses \texttt{0000} (its main
address), \texttt{0010}, \texttt{0001}, and \texttt{0011}.

The first parameter to fix is the dimension $d$ of the hypercube,
which is known {\em a priori} by all the participants of the
network. On the one hand, this parameter limits the maximum number
of nodes, but on the other hand, gives more flexibility to
connecting nodes due to secondary addresses. The problem is that
each new node should be adjacent to a maximum number of nodes,
ideally to all nodes within its radio coverage, in order to be
strongly connected. Intuitively, the larger the addressing space,
the richer the nodes' choice. We address this issue in detail in
Section~\ref{prac_cases_sty}.

\subsection{Indirect routing in the hypercube}~\label{addr_spa:ind_rout}

Recall that using an indirect routing technique means that there
are two phases for forwarding information: $(i)$ the source
asks, to the rendezvous node, the destination's address using its
universal identifier, $(ii)$ the source sends the messages to
the destination. This mechanism presupposes that there exists a
method to find the rendezvous node, because the only available
information is the destination's rendezvous address $V$ which is managed by a
certain node.

As previously seen, the main address and the addressing space are given
by already connected nodes. When a node gives an address, it also
delegates a portion of its used addressing space (generally the
upper half of it) to a new incoming node. For example, in
Fig.~\ref{routingtopo} the node {\tt 0000m2} would give the main
address and addressing space {\tt 0010m3} to a new node, causing
the change in the {\tt 0000} mask: from {\tt m2} to {\tt m3}, and
it sends all the address resolution information stored for this
addressing space. This means that the main address of a new node
is {\tt 0010}, and it manage the addresses {\tt 0010} and {\tt
0011}. The utilization of this method for all the nodes causes a
tree distribution of the network addresses, which we call ${\cal
T}$ in the remainder of this paper. 
Fig.~\ref{routingtopo} presents a real topology, where cutting the 
link between nodes {\tt 0111} and {\tt 1111} we can observe an example 
of the ${\cal T}$ tree.

Therefore, for a given rendezvous address $V$, we should find all the
possible nodes which can manage it in their addressing space. This
task might be very simple using the ${\cal T}$ tree.
In this case, it is enough to move through the tree following 
the match of the rendezvous address $V$'s prefix.
Again, this search is trivial for the
complete hypercube, but in an incomplete case one needs to find
the ${\cal T}$ tree. In a normal operation, ${\cal T}$ always
exists. We handle different cases in Section~\ref{prac_cases_sty}.

\section{Design issues: Proactive or reactive?}~\label{pro_reac}

We present two routing methods in this paper: proactive and
reactive. The first builds and maintains the routing tables all
the time, and assures a route for every node in a network. The
second method finds a route on demand, and maintains the route for
a given period of time. Clearly, the proactive approach is very
useful for quite stable networks, ({\em i.e.} where node mobility
is low and nodes' lifetime is long). For highly dynamic networks,
where nodes are joining and leaving all the time, 
the reactive method is more appropriate.

\subsection{Case 1: Proactive routing protocol}~\label{pro_reac:pro_rout}

In a complete hypercube, there is no problem for routing, because
all nodes and edges exist, then it is possible to use the
adjacency properties of the hypercube. In a general case, we
should propose a routing table composed of a combination of
default entries and some other routing entries. The default
entries take advantage of the address assignment method (the
${\cal T}$ tree). The other entries consist in a set of routes for
other connections which do not belong to ${\cal T}$, represented
by the secondary addresses. In other words, we put one entry in a
routing table for each connection of the node, and also for the
shortest advised routes.
Because the address assignment method, each node $v$ has a parent 
node and it may also has some children nodes,
noted by 


\begin{itemize}
\item {\em Parent node}: $P_v$ is the node that assigns
      a main address to node $v$. The parent node also delegates
	  a portion of its addressing space to node $v$.
\item {\em Child node}: $C^{i}_{v}$ is the node that has
      node $v$ as parent node, {\em i.e.} $P_{C^i_v} = v$, $1 \leq i \leq
      k$, being $k$ the number of $v$ children nodes.
\item {\em Children set}: represented by ${\bf C}_{v}~=
      \{C^{1}_v,\ldots,C^{k}_v\}$, is the set of children nodes.
\end{itemize}

The address assignment method is formalized as follows. The main
address of node $v$ is {\tt $p\_0\;$m$\;b$}, where $p$ is the
prefix of the $v$ address, $\_0$ is the zeros which completes the
address length, and $b$ is the number of bits from the left. The
prefix is obtained by doing $v\;{\tt AND}\;M_v$, where 
$M_v=\sum_{j=b}^{d-1}2^j$. 
Thus, the node $v$ assigns an address as following
\begin{equation}
p_v\_0\;{\tt m}\; b_v\; \xrightarrow{\tt address\_assignament}\;  \left\{
\parbox{4cm}{
       $p_v\_0\;{\tt m}\;(b_v+1)$ \\
       $p_v\_0+2^{d-b_v-1}\;{\tt m}\; (b_v+1) $
      }
\right.
\end{equation}

The parent node $P_v$ has always the main address $p_v\_0-2^{d-\beta_v}$, 
where $\beta_v$ is the first value of $b_v$, {\em i.e.} when the 
main address of $v$ was assigned. 
Each child $C^i_{v}$ in the children set ${\bf C}_{v}$, when they 
exist, has as main address
\mbox{$p_v\_0+2^{(d-1)-x_i}\quad, \forall x_i \in \{\beta_v,\beta_v+1,\cdots,d-1\}$}.
Note that the child index is defined as $i=x_i-\beta_v+1$.

Each entry in a routing table is composed of a  prefix, a mask,
and a next hop. The masks have the same form as in the IP case,
{\em i.e.} the number of ones from the left side.

As mentioned before, there are two types of entries:

\begin{itemize}
\item the entries of ${\cal T}$ tree, {\em e.g.}, \mbox{$0\_0/0\rightarrow
      p_v\_0-2^{d-b}$} for the parent node $P_v$, and
\mbox{$p_v\_0+2^{d-1-x_i}\;/\;x_i\rightarrow p_v\_0+2^{d-1-x_i}$} for each 
     child node $C^{x_i-\beta_v+1}_v$ ;
\item the entries for a neighbor $t$ ({\it i.e.} $w$, $u$, and $z$ in
      the example) which
      does not belong to the ${\cal T}$ tree is $p_t\_0/a_t^v \rightarrow
      t$, where $p_t\_0$ is the prefix obtained applying the mask defined
      by $a_t^v$, as $M_t=\sum_{j=a_t^v}^{d-1}2^j$.
\end{itemize}

The entries at $v$'s routing table are

\begin{center}
\begin{tabular}{|rccl|} \hline
$p_w\_0$ & $/\;a_w^v$ & $\rightarrow$ & $w$\\
$p_v\_0+2^{d-1-x_n}$ & $/\;x_n$ & $\rightarrow$ & $p_v\_0+2^{d-1-x_n}$\\
\vdots & \vdots & $\rightarrow$ & \vdots \\
$p_u\_0$ & $/\;a_u^v$ & $\rightarrow$ & $u$\\
$p_v\_0+2^{(d-1)-x_1}$ & $/\;x_1$ & $\rightarrow$ & $p_v\_0+2^{(d-1)-x_1}$\\
$p_z\_0$ & $/\;a_z^v$ & $\rightarrow$ & $z$\\
$0\_0$ & $/\;0$ & $\rightarrow$ & $p_v\_0-2^{d-b}$\\ \hline
\end{tabular}\\
\end{center}

\noindent where $a_w^v\geq x_n\geq\cdots\geq a_u^v \geq x_1 \geq a_z^v > 0$, 
and $x_i$ is the number of bits from the left, obtained after 
the $i^{\tt th}=x_i-\beta_v+1$ child ($C^i_{v}$ and $x_i \in 
\{\beta_v,\beta_v+1,\cdots,d-1\}$). 
The order is very important because the first matching is used for routing.

These entries are determined by Algorithm~\ref{algo1} when a local
node $v$ is connected to $u\notin {\bf C}_v$. 
The first step computes the node $y$ which is in the middle of the 
path from $v$ to $u$ in the tree ${\cal T}$.
Then, it computes $s$, which is the length of the matching prefix, 
either of $v$ or of $u$, because $y$ is ancestor o $v$ or $u$.    
Finally, a message advertising the new route
is sent to all neighbors. Then, once receiving the message each 
neighbor $u$ executes the Algorithm~\ref{algo2} to add and resend the new 
received routes when
necessary. In this algorithm, $d_H(\cdot,\cdot)$ is the distance
in the hypercube.

\begin{table}[h]
\begin{algorithm}{Routing tables construction at node $v$}\label{algo1}
\\
Reach a node $y$, such that $d(y,x)\leq d(v,y) \leq d(y,x)+1$, where
$d(\cdot,\cdot)$ is the distance on the default tree ${\cal T}$.
\\
Set the entry $y/s\texttt{->} x$ in $v$'s routing table, where
$s$ is the number of unchanged bits between $y$ and, $x$ if it is a $y$'s descendant in a ${\cal T}$, else $v$ is a $y$'s descendant. \\
Send a message to all neighbors, except $x$, with $y/s\texttt{->}
v$.
\end{algorithm}
\end{table}
\begin{table}[h]
\begin{algorithm}{Forwarding routing tables messages}\label{algo2}
\\
Node $u$ receives $\{y/s\texttt{->} v\}$ from neighbor $v$ \\
{\bf If} the $d(y,u) \leq d_H(y,v)+1$ {\bf then} \> \\
    Add the entry $y/s\texttt{->} v$ \\
    Send a message to all the neighbors, except $v$, with $\{y/s\texttt{->} u\}$.
    \<
\end{algorithm}
\end{table}

We should consider also the case when a node $v$ lost the connection with its parent node $P_v$.
In this case it sends a message ${\cal M}$ to its neighbors, in order to find a connection with the ${\cal T}$ tree.
This message ${\cal M}$ is resent by each node until one, {\em e.g.} $w$, which is connected to its parent node $P_w$ and the prefix $P_v$ of the first node $v$ is not contained in $P_w$.
Then, node $w$ resends a message reply to $v$ which confirms and sets the default route of $v$: $0/0 \rightarrow u$, such as $u$ is the $v$'s neighbor having a path to $w$.
The node $w$ also sends a message, following the ${\cal T}$ tree, to reach $P_v$ or its closer ancestor, we call this node ${\cal P}_v$.
The objective is to establish a route from ${\cal P}_v$ to $v$ passing by $w$, restoring the ${\cal T}$ tree.
In this way the ${\cal T}$ tree is reconnected, assuring the default route for nodes $v$ and $C_v^i$.

\subsection{Case 2: Reactive routing protocol}~\label{pro_reac:reac_rout}

In our case, the logical topology is built following adjacent
addresses, hence there is a coherent mapping between the physical
positions and the logical addresses.

There are two complementary methods for routing: the first is for
address resolution messages and the second is for other messages.

\begin{figure}
\begin{center}
\includegraphics[width=45mm,angle=0]{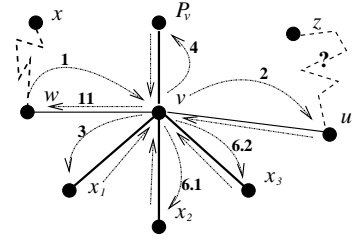}
\end{center}
\caption{Execution of Algorithm~\ref{algo3}} \label{algo3_example}
\end{figure}

Let us begin with the second case. This method considers that the
hypercube is complete, and routes the message by sending it to
neighbors whose addresses are closer to the destination. When a
message is blocked, {\em i.e.} there is no route, the message goes
backwards and it is sent through a different route, leaving a mark
on the unsuccessful route. Algorithm~\ref{algo3} presents the
method used to forward a message at node $v$, received of node $w$, when the source is
$x$ and the destination is $z$.
Fig.~\ref{algo3_example} shows an example where there is no route from $v$ to $z$.
The number over the arrows corresponds to step number of the algorithm. 
The curved arrows are the sent message ${\cal M}$ and the right arrows is the return of the message ${\cal M}$.
The special case of arrows with $6.1$ and $6.2$ correspond to the first and second iteration of the loop, respectively.

\begin{table}[h]
\begin{algorithm}{Forwarding in reactive routing at node $v$}\label{algo3}
\\
$v$ receives a message ${\cal M}(x,z)$ form neighbor $w$.\\
$v$ sends the message to a neighbor $u \neq \{P_v,w\}$, such that $u$ minimizes $d_H(u,z)$ \label{algo3:2}\\
{\bf If} there is no route from $u$ {\bf then} mark this route and resend the message to other neighbor $\neq \{P_v,w\}$. \label{algo3:3}\\
{\bf If} the message is returned again {\bf then} send the message to its parent $P_v$ in ${\cal T}$ and mark all the remaining neighbors as unexplored. \\
{\bf If} the message returns {\bf then} do \> 
      {\bf until} all neighbors are explored:~\label{loop} \> \\
          send the message to a neighbor marked as unexplored  \\
          {\bf if} the message returns {\bf then} mark this neighbor as blocked, and return to step~\ref{loop}. \< \<\\
{\bf If} there is no route {\bf then} \> \\
   {\bf If} the original sender is the local node $v$: \> \\
        {\bf then} ${no\_route\_to\_host := true}$. \\
	{\bf else } resend the message ${\cal M}(x,z)$ to the neighbor sender $w$. \< \< 
\end{algorithm}
\end{table}

Remember that $d_H(\cdot,\cdot)$ is the distance in the hypercube,
and ${\cal T}$ is the initial tree used for distributing the
addressing space. This algorithm favors the exploration of 
farther regions from the root of ${\cal T}$. If it does not
find a route then it sends the message towards the root, and
finally if it still does not find a route, it performs an
exhaustive exploration. A timer is used by resetting the marks in
unsuccessful routes, but they can also reset by an update message.
The value of this timer is long, and is only used to give a robust
behavior, {\em i.e.} when an update message is lost.

The update messages are sent when new topological connections are made.
When a node $v$ has been connected with another node $w$,
node $v$ sends update messages with its address and the new neighbor 
address $w$ to all its neighbors ($w$ does not consider this message). 
Other case is when $v$ receives an update message from a neighbor $u$, then $v$ clears the blocked routes in the $u$ corresponding interface.

The first routing case, which corresponds to a resolution request, 
uses a variation of Algorithm~\ref{algo3}.
This variation consist in, firstly to change of step~\ref{algo3:2}, 
and secondly to eliminate the step~\ref{algo3:3}.
The elimination of step~\ref{algo3:3} is motivated to give more priority, 
to address resolution messages, to reach their destination. 
It is clear that the number of address resolution messages
\footnote{Discovered addresses are stored
in a local cache table and associated to a timeout. Resolution messages
are sent one time for the first communication, and then, when the 
timeout of the corresponding cache table's entry has expired.} are 
lower than the data messages,
and then they have less contribution to the 
congestion of the ${\cal T}$'s root. 
The step~\ref{algo3:2} of Algorithm~\ref{algo3} is replaced by

\vspace{2mm}
\small
\noindent
\begin{tabular}{cc}
\hline
\ref{algo3:2} & $v$ sends the message to a neighbor $u \neq w$, such that \\
 & $u$ minimizes $d_{\cal T}(u,p_z\_0)\; \forall \; s \;/\; p_z\_0 = z \;{\tt AND} \sum_{i=s}^{d-1} 2^i$. 
\vspace{1mm} \\
\hline
\end{tabular}
\normalsize
\vspace{2mm}

That is, it finds the neighbor which minimizes the distant to one of the possible prefixes of the virtual address in the ${\cal T}$ tree.
The reason is that the virtual address is contained in the managed addressing space of a certain node, because the ${\cal T}$ tree distribution method. 

%

\section{Practical considerations and case studies}\label{prac_cases_sty}

In this section we will consider the application of our
architecture in different scenarios. Then, we present two examples
for each routing method.

\subsection{Choosing the dimension $d$}~\label{prac_cases_sty:dim}

One important issue of hypercubes is the addressing space, because it
defines multiple possibilities of connection and routing. 
We consider two cases: sparse and dense networks. 
Given a fixed $d$, nodes are connected until their radio neighbors 
have not any available addresses.
In sparse case, nodes are mainly connected augmenting the diameter of 
the logical graph.
Dense networks, however, are susceptible to have a lot of connections 
per node, increasing the number of secondary addresses, consuming a 
lot of address per node, and given a small diameter of the logical graph.
Therefore, there is a trade-off between the radio coverage and the
maximum size of the network for choosing the dimension.

More precisely, the extreme case on sparse network is when a node
has only two neighbors, this results in  a linear chain with $2d$
nodes because the address distribution method follows a ${\cal T}$ tree. 
In general, the maximum number of nodes $n_{\max}$ that can join
a sparse network with $k$ neighbors is
\begin{equation*}
n_{\max}=\sum_{i=1}^{k} s(d-i,k)  \;, \quad  \forall\; 2 < k < d
\end{equation*}
where $d$ is the dimension of the hypercube, and $s(\cdot,\cdot)$ is 
the following recursive function
\begin{equation*}
s(h,k)=\left\{
  \begin{array}{ll}
        \sum_{j=1}^{k-1}  s(h-j,k)\;,\; & \forall\; h  > k\\
         2^h                      \;,\; & \forall\; h \leq k \\
  \end{array}\right.
\end{equation*}

For dense networks, the number of addresses in each node depends
on the number of physical neighbors, considering that all nodes
could be obtained from a compatible secondary address with their
neighbors. Therefore, a high percentage of neighbors of a node are
connected among them, which means that the network has a lot of
triangles. If the percentage is denoted by $c<1$, $k$ is the
number of neighbors, and $d$ is the number of dimensions, then, for
each $c\cdot k$ nodes there is a clique\footnote{In a clique of 
$n$ nodes each node
is connected to all nodes, and the total number of connections is
$n(n-1)/2$.}. Consequently, if $n_{\max}$ is the number of nodes 
that can join a dense network, there are $n_{\max}/(c\cdot k)$ cliques  
and $\frac{c\cdot k (c\cdot k-1)}{2}$ number of connections 
, {\em i.e.} secondary addresses, for each clique. Then,
\begin{eqnarray*}
 \frac{n_{\max}}{c\cdot k} \cdot\frac{c\cdot k (c\cdot k-1)}{2} &\leq& 2^d \nonumber \\
 n_{\max}(c\cdot k-1) &\leq& 2^{d+1} \nonumber \\
 n_{\max} &\leq& \frac{2^{d+1}}{c\cdot k-1} \enspace,
\end{eqnarray*}
where $2^d$ is the total number of nodes in a $d$-dimensional hypercube.

A useful approximation of maximum path length, for both cases, is the following.
Considering $n(\ell)$ the number of total neighbors up-to distance $\ell$ 
for a node in a $k$ regular network ({\em i.e.}, each node has $k$ neighbors).
Then, for $\ell=2$ we have $n(\ell)=(k-1)^2+1$, because the neighbors at distance 
$1$ are $k$, and each of these neighbors has other $k-1$ different neighbors.
The maximum path length $\ell_{\max}$ for a network with $n$ nodes is  
\begin{eqnarray*}
n & = & n(\ell_{\max}) \\
n & = & (k-1)^{\ell_{\max}} + 1\\
\log_{k-1} n &\simeq & \ell_{\max}  \enspace, \\
\end{eqnarray*}
which is valid for $k<d/2$.
The main difference of $\ell_{\max}$ between sparse and dense networks is 
the value of $k$, because dense networks has a higher $k$ than sparse ones, 
thus the maximum path distance will be smaller in dense networks.

Therefore, considering the general purpose case, where the
addresses are not too long and where it is also possible to obtain
some secondary addresses, an empirical choice of $d$ could be
$n_{\max}=2^{4d/5}$. That is, we propose to increase the
addressing space by 20\% of the address length, allowing up to
$2^{d/5}$ secondary addresses per node.

\subsection{An example of the proactive protocol}~\label{prac_cases_sty:pro}

We present here examples of the routing table construction,
communication between two nodes, and address resolution.

For the proactive method, each node has a pre-established table.
Consider Fig.~\ref{routingtopo} and the routing table of node {\tt
1000m3}: \hfill

{\small
\begin{center}
\begin{tabular}{|c|c|}
\hline destination & next hop \\ \hline
{\tt 1010/3} & {\tt -> 1010} \\
{\tt 1100/2} & {\tt -> 1100} \\
{\tt 0000/0} & {\tt -> 0000} \\
\hline
\end{tabular}
\end{center}
}

The first entry means that all messages addressed to destinations
whose most significant bits are {\tt 101} must be sent through
node {\tt 1010} (one of its children). The second line is for
addresses attained through the child {\tt 1100}. It is worth
noting here the strict relationship between the addressing space
of a child and the destination entry in the table at the time the
child was connected, {\em e.g.}, the entry {\tt 1100/2} and its
first child {\tt 1100m2}. Currently, node {\tt 1100} has mask {\tt
m3} because it has already assigned an address to a new node (but
its mask was {\tt m2} before the arrival of the new node). We call
the addressing space of a node at the time it joins the network
the {\em initial addressing space} of the node.

Finally, the last line is the default route to its
parent node {\tt 0000}. (Note that ``{\tt /0}'' means the first
``0'' most significant bits.) The default route is represented by
{\tt 0000/0} because it matches all nodes.

It is important to stress that the order of the lines in the
routing table is important. The first line is the most
constraining entry, because the 3rd most significant bits must
match (due to ``{\tt /3}''). The last line is the least
constraining entry, hence, the default route entry. The first node
in the network does not have a default route, because it has no
parent and it is the parent of all nodes. However, it has entries
for its children, then all the possible addresses in the hypercube
are represented.

There are others types of entries in order to represent a
connection that does not follow the tree structure. This is the
goal of our proposal. For example, Fig.~\ref{routingtopo} displays
the connection between nodes {\tt 1111} and {\tt 0111}, and the
corresponding routing tables. In this scenario, node {\tt 1111}
has the following routing table:

{\small
\begin{center}
\begin{tabular}{|c|c|}
\hline destination & next hop \\ \hline
{\tt 0000/1} & {\tt -> 0111} \\
{\tt 0000/0} & {\tt -> 1110} \\
\hline
\end{tabular}
\end{center}
}

The default route is through the node's parent, and the other
route means that all the addresses whose most significant bit is
{\tt 0} can be reached through node {\tt 0111}. This entry, at
local node $v=${\tt 1111}, can be determined by
Algorithm~\ref{algo1} after the connection with $u=${\tt 0111}.

Now we illustrate a case where a node exchanges data. Consider
that node \texttt{1110} sends a message to node \texttt{0110}. The
first entry in the routing table of \texttt{1110} is
\texttt{1111/4 ->1111}. This means that the comparison is done
using the four most significant bits (because of ``\texttt{/4}'')
of the destination node \texttt{0110}. We observe that
the final destination is different to the entry at routing table, 
{\em i.e.} \texttt{0110}$\neq$\texttt{1111}, 
and therefore the matching fails.
The second line is \texttt{0100/2 ->1111}, the two most
significant bits of the destination are \texttt{01}, and they
equal the two most significant bits of \texttt{0100/2}. Therefore,
this entry matches and the packet is forwarded to node
\texttt{1111}. The first entry of the routing table of
\texttt{1111} is \texttt{0000/1 ->0111} and the most significant
bit of destination is \texttt{0}~-- this entry matches and the
packet is forwarded to \texttt{0111}. As \texttt{0111} is a
secondary address, the packet is now at node \texttt{0110}, which
is the final destination address.

Finally, we present an address resolution request. This kind of
message is routed in the same form as data messages. The only
difference is that the destination, {\em i.e.} the rendezvous
address, may or may not be the main address of a node. If it is
not the main address, the message will arrive at the node which
manages this address. Therefore, before applying the routing
algorithm, each node must verify if the destination belongs to
addresses that it manages. For example, node \texttt{0110} wants
to know which is the network address of a particular identifier
$U$. Then it applies the hash function to know the rendezvous
address, that is $hash(U)=$\texttt{1101}. Because this address is
not managed by the local node \texttt{0110m3}, it sends the
message to \texttt{1101}. The first entry in \texttt{0110}'s
routing table is \texttt{1100/2 ->1111}, and it matches because
the two most significant bits of \texttt{1101} are \texttt{11}.
Then the request message is sent to node \texttt{1111}. This node
does not manage the address in the request either, so it forwards
the message using its routing table. The first entry is
\texttt{0000/1 ->0111}, which does not match. The second is
\texttt{0000/0 ->1110}, which matches because it is the default
routing entry, and the message is forwarded to node \texttt{1110}.
Since this node has a \texttt{m4} mask, it does not manage the
address into the request, so it will forward the message. The
first entry in its routing table is \texttt{1111/4 ->1111}, which
does not match, and the second one is \texttt{0100/2 ->1111} which
does not match either. Finally, the last entry matches because it
is the default route. The node \texttt{1100} receives the request
for the server resolution of address \texttt{1101}, and the
addresses managed by \texttt{1100m3} are \texttt{1100} and
\texttt{1101}. This node looks up the network address $E$ corresponding to node $U$,
and sends a reply to the source node {\tt 0110} with the network
address $E$. The source can then directly communicate to the node
whose address is $E$.

\subsection{An example of the reactive protocol}~\label{prac_cases_sty:reac}

In the reactive case, there are no routing tables, but some
information concerning temporary path recently used by each node.
This information is created in a communication step, storing the
unsuccessful paths. In this section we present two communication
cases and an address resolution procedure.

Because this method starts with no {\em a priori} knowledge of how
complete the hypercube is, it uses standard routing in hypercubes.
This means that routing is done by changing the different bits one
by one, {\em i.e.} sending to neighbors closer to the destination
(recall that a node is a neighbor if their addresses differ on one
bit). For example, if node \texttt{0100} sends a message to
\texttt{1111}, it does \mbox{(\texttt{0100} XOR
\texttt{1111})=\texttt{1011},} that is the first, third, and
fourth bits change. Then node {\tt 0100} can send the message to
one of the following neighbors: \texttt{1100}, \texttt{0110} or
\texttt{0101}, because they differ, from {\tt 0100}, in only one
bit. The only node present in the network is \texttt{0110} (see
Fig.~\ref{routingtopo}), therefore the message is forwarded to
this node. At node \texttt{0110}, XOR is applied again, which
results in \texttt{1001}. The only existing neighbor is
\texttt{0111}, which corresponds its secondary address. Finally,
the result of XOR is \texttt{0001}, and the neighbor \texttt{1111}
is the last step.

We illustrate a more complicated case with the following example.
Node \texttt{1000} sends a message to node \texttt{0110}, then
\mbox{(\texttt{1000} XOR \texttt{0110}) = \texttt{1110}}, and the
possible forwarders in the network are \texttt{1010} and
\texttt{1100}. Node \texttt{1000} sends then the message through
\texttt{1010}. Candidate forwarder neighbors of node \texttt{1010}
are \texttt{1110} and \texttt{0010}, because \mbox{(\texttt{1010}
XOR \texttt{0110}) = \texttt{1100}}. But \texttt{0010} does not
exist in the network and \texttt{1110} is not connected to it.
Node \texttt{1010} sends the message backwards, and node
\texttt{1000} sets a temporary entry because now it knows that
there exists no path. Of course, this entry should be removed
after a timeout, or if the node becomes connected to other nodes.
Finally, the message is forwarded to node {\tt 1100}. At this
node, the result of \mbox{({\tt 1100} XOR {\tt 0110})} is {\tt
1010}, then a possible forwarder, present in the network, is {\tt
1110}. This latter receives the message and computes \mbox{({\tt
1110} XOR {\tt 0110}) = {\tt 1000}}, but the nodes {\tt 1110} and
{\tt 0110} are not interconnected. In this case, it is better to
take a new path in the opposite way. Then, the message is sent to
node {\tt 1111}. This node computes \mbox{({\tt 1111} XOR {\tt
0110}) = {\tt 1001}}, and the possible forwarder is {\tt 0111}. As
{\tt 0111} is a secondary address and its primary address is {\tt
0110}, the message has arrived to the final destination.

For the address resolution case, we use the modified Algorithm~\ref{algo3}.
Suppose that node {\tt 1110} wants to send a message to  node with
universal address $U$, then it obtains  $hash(U)=${\tt 0101} (the
rendezvous address). 
The node who minimize $d_{\cal T}({\tt 1110},{\tt 0101})$ is its parent node $P_v={\tt 1100}$. 
Since the other nodes are in a
similar situation, the message is forwarded to consecutive parent
nodes until it reaches {\tt 0000}. Because the first most
significant bit is the same as the desired address {\tt 0101}, the
actual node checks if this address belongs to its managed space.
The result is negative and the message is sent to the neighbor
{\tt 0100} which is the closest to {\tt 0101}. This node has in
its managed space the addresses {\tt 0100} and {\tt 0101} (because
its mask is {\tt m3}). Therefore, node {\tt 0100} looks up the
virtual address and sends it to {\tt 1110} in a response message.
In this case, the communication was done using the ${\cal T}$
tree. If the ${\cal T}$ tree is disconnected,  
the message is sent backwards until a route
is found, as in the data communication case.

\section{Discussion and further research}~\label{discuss}

The most effective protocol to self-organization networks is a 
combination of a good physical-to-logical mapping with a simple 
and robust routing protocol, and small routing tables.  The 
geographical routing could be the most promising, but the 
reception of GPS can not be enough, {\em e.g.},  inside of a building. 
Moreover, the GPS error, which depends also the reception quality, 
is too large for some dense networks. Next candidates are those 
that use indirect routing and build a logical and mathematical 
structure from mere connectivity between nodes.  Up to now, this 
protocols propose a logical tree for connecting nodes~\cite{benjie02,
eriksson04,viana.winet04,eriksson03,tsuchiya88,tsuchiya87,viana_percom03}. 

In the deployment of self-organized systems, flexibility in route 
selection is an important issue to be considered, which affects 
the performance in terms of path length,
traffic concentration, and resilience to failures. In this
context, the organization of the addressing structure has
a strong influence. In the tree-based structures, paths are 
limited by the hierarchical structure of a
tree~-- there is only one path between any two nodes.
A tree offers low flexibility in
route selection, contrary to the greater flexibility offered by
the multi-dimensional approaches.
Our hypercube approach offers multiple links 
options that get the path closer to the physical distance. 

A spontaneous network could have a well balanced traffic only when the distance between two nodes is closer to their physical distance.
In a case that the logical structure is a tree, is very difficult to fill this condition, mainly because the connection order. 
Even, following the optimal connection order, when the density of nodes is high, a message sent to a physical neighbor should pass to other node before to arrive at the destination.  
Instead, the incomplete hypercube is better because 
it allows multiple links, even for far nodes, giving more 
privilege to the neighbors connections. This also makes a more 
coherent physical-to-logical mapping, given a similar physical 
and logical distances. Therefore, using the hypercube as 
underling logical structure, coupled with indirect routing, 
we provide redundant connections, a better load distribution
and two different routing methods. These characteristics permit 
to cover a wide range of applications according to their 
mobility characteristics.

Although the  greater flexibility in route selection offered by
the multi-dimensional approaches, their associated addressing and
location models are more complex, contrarily to simple manageable
structures offered by tree-like structures.
The main problem with the incomplete hypercubes could be their 
relative complexity, but evidently there exist a trade-off 
between the simplicity and the robustness. Our proposal provides 
the advantages of a good physical-to-logical mapping, and two 
routing protocols adapted to the mobility characteristics, 
given a robust behavior.

Since this paper had as objective to describe a hypercube-based
architecture to implement indirect routing, future research includes
a complete evaluation of the proposed protocol under fixed and
mobile environments. 
Some optimization mechanisms and implementation issues for
improving robustness in terms of location information
availability, load balancing, and failures are also interesting to
analyze.

{\small

\bibliographystyle{ieeetr-lui}
\bibliography{bib28-01-2005}

\begin{thebibliography}{10}

\bibitem{ratnasamy01}
S.~Ratnasamy, P.~Francis, M.~Handley, R.~Karp, and S.~Shenker, ``A scalable
  content-addressable network,'' in {\em Proceedings of {ACM SIGCOMM'01}}, Aug.
  2001.

\bibitem{stoica.tnet03}
I.~Stoica, R.~Morris, D.~Liben-Nowell, D.~R. Karger, M.~F. Kaashoek, F.~Dabek,
  and H.~Balakrishnan, ``Chord: a scalable peer-to-peer lookup protocol for
  internet applications,'' {\em IEEE/ACM Transactions on Networking}, vol.~11,
  no.~1, pp.~17--32, Feb. 2003.

\bibitem{rowstron01}
A.~Rowstron and P.~Druschel, ``Pastry: Scalable, distributed object location
  and routing for large-scale peer-to-peer systems,'' in {\em Proceedings of
  {IFIP/ACM Middleware'01}}, Nov. 2001.

\bibitem{blazevic01}
L.~Blazevic, L.~Buttyan, S.~G. S.~Capkun, J.~P. Hubaux, and J.~Y.~L. Boudec,
  ``Self-organization in mobile ad-hoc networks: the approach of terminodes,''
  {\em IEEE Computer Communications Magazine}, June 2001.

\bibitem{jinyangli00}
J.~Li, J.~Jannotti, D.~S. J.~D. Couto, D.~R. Karger, and R.~Morris, ``A
  scalable location service for geographic ad hoc routing,'' in {\em
  Proceedings of {ACM MOBICOM'00}}, Aug. 2000.

\bibitem{xue01}
Y.~Xue, B.~Li, and K.~Nahrstedt, ``A scalable location management scheme in
  mobile ad-hoc networks,'' in {\em Proceedings of {IEEE Conference on Local
  Computer Networks (LCN)}}, (Tampa, FL, USA), Nov. 2001.

\bibitem{benjie02}
B.~Chen and R.~Morris, ``L+: Scalable landmark routing and address lookup for
  multi-hop wireless networks,'' tech. rep., Massachusetts Institute of
  Technology, Cambridge, Massachusetts - MIT LCS Technical Report 837
  (MIT-LCS-TR-837), Mar. 2002.

\bibitem{eriksson04}
J.~Eriksson, M.~Faloutsos, and S.~Krishnamurthy, ``Scalable ad hoc routing: The
  case for dynamic addressing,'' in {\em Proceedings of {IEEE INFOCOM'04}},
  (Hong Kong), Mar. 2004.

\bibitem{viana.winet04}
A.~C. Viana, M.~D. Amorim, S.~Fdida, and J.~F. Rezende, ``Indirect routing
  using distributed location information,'' {\em ACM Wireless Networks},
  vol.~10, no.~6, pp.~747--758, Dec. 2004.

\bibitem{viana.adhoc05}
A.~C. Viana, M.~D. Amorim, S.~Fdida, and J.~F. Rezende, ``Self-organization in
  spontaneous networks: the approach of dht-based routing protocols.'' to
  appear in Ad Hoc Networks Journal, 2005.

\bibitem{hubaux01}
J.~P. Hubaux, T.~Gross, J.~Y.~L. Boudec, and M.~Vetterli, ``Towards
  self-organized mobile ad hoc networks: the terminodes project,'' {\em IEEE
  Communications Magazine}, vol.~39, no.~1, pp.~118--124, Jan. 2001.

\bibitem{terminode}
{Terminodes Project}.
\newblock {\tt http://www.terminodes.com/}.

\bibitem{grid}
{Grid Project}.
\newblock {\tt http://www.pdos.lcs.mit.edu/grid/}.

\bibitem{eriksson03}
J.~Eriksson, M.~Faloutsos, and S.~Krishnamurthy, ``Peernet: Pushing
  peer-to-peer down the stack,'' {\em Proceedings of {International Workshop on
  Peer-To-Peer Systems (IPTPS'03)}}, Feb. 2003.

\bibitem{tsuchiya88}
P.~F. Tsuchiya, ``The landmark hierarchy: a new hierarchy for routing in very
  large networks,'' in {\em Proceedings of {ACM SIGCOMM'88}}, Aug. 1988.

\bibitem{tsuchiya87}
P.~F. Tsuchiya, ``Landmark routing: Architecture, algorithms and issues,''
  tech. rep., MTR-87W00174, MITRE Corporation, Sept. 1987.

\bibitem{broch98}
J.~Broch, D.~A. Maltz, D.~B. Johnson, Y.~Hu, and J.~Jetcheva, ``A performance
  comparison of multi-hop wireless ad hoc network routing protocols,'' in {\em
  Proceedings of {ACM MOBICOM'98}}, Oct. 1998.

\bibitem{tseng99}
S.~Ni, Y.~Tseng, Y.~Chen, and J.~Sheu, ``The broadcast storm problem in a
  mobile ad hoc network,'' in {\em Proceedings of {ACM MOBICOM'99}},
  pp.~152--162, Aug. 1999.

\bibitem{stoica02}
I.~Stoica, D.~Adkins, S.~Zhuang, S.~Shenker, and S.~Surana, ``Internet
  indirection infrastructure,'' in {\em Proceedings of {ACM SIGCOMM'02}}, Aug.
  2002.

\bibitem{perkins94}
C.~E. Perkins and P.~Bhagwat, ``Highly dynamic destination sequenced
  distance-vector routing (dsdv) for mobile computers,'' in {\em Proceedings of
  {ACM SIGCOMM'94}}, Oct. 1994.

\bibitem{Leighton91}
F.~T. Leighton, {\em Introduction to parallel algorithms and architectures:
  array, trees, hypercubes}.
\newblock Morgan Kaufmann Publishers Inc. San Francisco, CA, US, 1991.

\bibitem{Euseuk01}
E.~Oh and J.~Chen, ``Parallel routing in hypercube networks with faulty
  nodes,'' in {\em IEEE International Conference on Parallel and Distributed
  Systems (ICPADS '01)}, pp.~338--345, July 2001.

\bibitem{scholsserAl03}
M.~Schlosser, M.~Sintek, S.~Decker, and W.~Nejdl, ``Hypercup - hypercubes,
  ontologies, and efficient search on peer-to-peer networks,'' in {\em Agents
  and Peer-to-Peer Computing: A Promising Combination of Paradigms, LNCS 2530},
  pp.~112--124, July 2003.

\bibitem{jimenez94}
M.~A. Jimenez-Montano, C.~R. de~la Mora-Basanez, and T.~Poeschel, ``On the
  hypercube structure of the genetic code,'' in {\em Proceedings of
  Bioinformatics and Genome Research}, pp.~445--459, Oct. 1994.

\bibitem{dajin_wang01}
D.~Wang, ``A low-cost fault-tolerant structure for the hypercube,'' {\em
  Journal of Supercomputing}, vol.~20, no.~3, Nov. 2001.

\bibitem{friedman99}
R.~Friedman, S.~Manor, and K.~Guo, ``Scalable stability detection using logical
  hypercube,'' tech. rep., Technion, Department of Computer Science Technical
  Report 0960, May 1999.

\bibitem{slack03}
J.~Slack, ``Visualization of embedded binary trees in the hypercube,'' tech.
  rep., Final Report of the Project for Information Visualization, Department
  of Computer Science, University of British Columbia, Apr. 2003.

\bibitem{saad85}
Y.~Saad, ``Data communication in hypercubes,'' tech. rep., Research Report 428,
  Department of Computer Science, Yale University, New Haven, CT, 1985.

\bibitem{viana_percom03}
A.~C. Viana, M.~D. Amorim, S.~Fdida, and J.~F. Rezende, ``Indirect routing
  using distributed location information,'' in {\em Proceedings of {IEEE
  International Conference on Pervasive Computing and Communications
  (PERCOM'03)}}, (Dallas-Fort Worth, Texas), Mar. 2003.

\end{thebibliography}

}

\end{document}